\documentclass[aps,prl,notitlepage,twocolumn,superscriptaddress]{revtex4-1}
\usepackage[utf8]{inputenc}
\usepackage[english]{babel}
\usepackage{amsmath}
\usepackage{amsfonts}
\usepackage{amssymb}
\usepackage{listings}
\usepackage{lipsum}
\usepackage{multirow}
\usepackage{datetime}
\usepackage{graphicx}
\usepackage{mathtools}
\usepackage{mathrsfs}
\usepackage{dcolumn}
\usepackage{multirow}
\usepackage{color}   
\usepackage{hyperref}
\hypersetup{
    colorlinks=true, 
    pdfborder = {0 0 0.5 [3 3]},
    anchorcolor=black,
    citecolor=blue,
    linktoc=all,    
    linktocpage=true,
    linkcolor=red,
	urlcolor=blue
}

\begin{document}

\title{Nonlinear treatment of a black hole mimicker ringdown}
\author{Nils Siemonsen}
\email[]{nils.siemonsen@princeton.edu}
\affiliation{Princeton Gravity Initiative, Princeton University, Princeton NJ 08544, USA}
\affiliation{Department of Physics, Princeton University, Princeton, NJ 08544, USA}

\date{\today}

\begin{abstract} 
We perform the first nonlinear and self-consistent study of the merger and ringdown of a black hole mimicking object with stable light rings. To that end, we numerically solve the full Einstein-Klein-Gordon equations governing the head-on collisions of a series of binary boson stars in the large-mass-ratio regime resulting in spinning horizonless remnants with stable light rings. We broadly confirm the appearance of features in the extracted gravitational waveforms expected based on perturbative methods: the signal from the prompt response of the remnants approaches that of a Kerr black hole in the large-compactness limit, and the subsequent emissions contain periodically appearing bursts akin to so-called gravitational wave echoes. However, these bursts occur at high frequencies and are sourced by perturbations of the remnant's internal degrees of freedom. Furthermore, the emitted waveforms also contain a large-amplitude and long-lived component comparable in frequency to black hole quasi-normal modes. We further characterize the emissions, obtain basic scaling relations of relevant timescales, and compute the energy emitted in gravitational waves.
\end{abstract}

\maketitle

\textit{Introduction} -- The black hole is a remarkably successful paradigm explaining astrophysical observations driven by highly compact and dark objects. Despite this success a large class of alternative objects has been developed \cite{Cardoso:2019rvt}, challenging this paradigm. These objects---black hole mimickers---are horizonless and imitate many or all of the observable signatures of black holes. Cardoso et al. pointed out that the light ring structure around these objects plays a central role in the ringdown of any black hole mimicker \cite{Cardoso:2016rao,Cardoso:2016oxy,Cardoso:2017cqb}. They observed that the gravitational wave (GW) emissions promptly after an extreme-mass-ratio merger of a binary black hole mimicker is universally identical to that of a black hole ringdown, independently of the mimicker's internal structure. The emitted signal deviates from the black hole ringdown only after a light-crossing time of the interior of the mimicker and is characterized by repeated burst-like GW echoes. As argued in Refs.~\cite{Cardoso:2016rao,Cardoso:2016oxy,Cardoso:2017cqb}, the prompt response can be understood as the partial reflection of the GWs sourced by the plunging companion off of the unstable light ring of the primary black hole mimicker. The subsequent echoes are due to repeated leakage of the transmitted gravitational perturbations traversing across the stable light ring in the mimicker's interior.

As a mechanism to discover as of yet unknown new physics, the ringdown of black hole mimickers has received much attention. Broadly, efforts have been devoted to understanding the ringdown of various classes of black hole mimickers \cite{Cardoso:2016rao,Cardoso:2016oxy,Holdom:2016nek,Bueno:2017hyj,Barcelo:2017lnx,Urbano:2018nrs,Raposo:2018rjn,Pani:2018flj,Cardoso:2019apo,Oshita:2018fqu,Wang:2019rcf,Oshita:2019sat,Maggio:2020jml,Dey:2020lhq,Ikeda:2021uvc}, computing waveforms and templates \cite{Mark:2017dnq,Nakano:2017fvh,Wang:2018gin,Burgess:2018pmm,Correia:2018apm,LongoMicchi:2019wsh,Maggio:2019zyv,Srivastava:2021uku,Annulli:2021ccn,Xin:2021zir,Ma:2022xmp}, determining detection prospects with GW detectors \cite{Maselli:2017tfq,Testa:2018bzd,Tsang:2018uie,LongoMicchi:2020cwm}, and searching for burst-like emissions in the GW data following known binary coalescence events \cite{Abedi:2016hgu,Ashton:2016xff,Westerweck:2017hus,Conklin:2017lwb,Nielsen:2018lkf,Lo:2018sep,Tsang:2019zra,Uchikata:2019frs,LIGOScientific:2020tif,LIGOScientific:2021sio,Uchikata:2023zcu,Miani:2023mgl} (see Ref.~\cite{Abedi:2020ujo} for a review). Despite the immense progress of this program, results were largely obtained modeling the internal structure of the mimicker by boundary conditions a small distance away from the would-be horizon and treating it's dynamical response at the test-field and extreme-mass-ratio level. Thereby neglecting, (i) nonlinear gravitational effects, (ii) coupling of the mimicker's internal structure to gravitational degrees of freedom even at the linear level, (iii) self-interactions of the matter making up the object, and (iv) any finite size effects of the objects. Notably, a few approaches have been developed in Refs.~\cite{Carballo-Rubio:2018jzw,Danielsson:2021ykm,Vellucci:2022hpl,Dailey:2023mvn} to address some of these shortcomings. However, a fully nonlinear and self-consistent treatment of the merger and ringdown of a black hole mimicker is still lacking. This leaves many important questions unanswered. In particular, when including all effects (i)--(iv), are the prompt emissions indeed identical to those of a ringing black hole? What are the amplitudes of the quasi-normal modes of the remnant excited during the merger? Specifically, is the waveform following the prompt ringdown solely characterized by GW echoes? How does the remnant's spin impact these conclusions?

In this work, we address these questions fully self-consistently in the large-mass-ratio and head-on merger setting. To that end, we drop all restrictions (i)-(iv) by performing a series of four fully nonlinear numerical time-domain evolutions of coalescing spinning binary boson stars---a particular set of black hole mimickers---resulting in spinning horizonless remnants with stable light rings. We find that the prompt dynamical response of the remnants, encoded in the emitted GWs, approaches that of Kerr black holes in the large-compactness limit. The subsequent emissions contain both a high-frequency burst-like component with frequency set the binary's size-ratio, and a large-amplitude and long-lived component indicative of excited trapped modes in the remnant's interiors.

\begin{table}[b]
\begin{ruledtabular}
\begin{tabular}{cccc|cc}
$C$ & $\tilde{\omega}/\mu$ & $J/M^2$ & $M\mu$ & $q$ & $\eta$ \\ \hline
$0.38$ & $0.105$ & $0.95$ & $18.7$ & $78$ & $4.4$\\
$0.36$ & $0.111$ & $0.98$ & $17.5$ & $73$ & $4.3$\\
$0.32$ & $0.130$ & $1.08$ & $13.9$ & $58$ & $3.9$\\
$0.28$ & $0.150$ & $1.22$ & $11.0$ & $46$ & $3.5$\\
\end{tabular}
\end{ruledtabular}
\caption{Properties of the primary boson star in the sequence of binaries. The secondary with frequency $\tilde{\omega}_*/\mu=0.85$, mass $M_*\mu=0.24$, and angular momentum $J_*/M_*^2=12.1$, is the same in all binaries considered. The last two columns show the mass- and size-ratios $q$ and $\eta$ of the binaries, respectively.}
\label{tab:propMain}
\end{table}

\textit{Model \& Methods} -- Boson stars \cite{Kaup:1968zz,Ruffini:1969qy} are the most developed highly compact and black hole mimicking objects that can be treated within numerical relativity \cite{Schunck:2003kk,Liebling:2012fv}. Those stars relevant for this work are solutions in the theory with Lagrangian \cite{Friedberg:1976me}
\begin{align}
\mathcal{L}=\frac{R}{16\pi}& -g^{\mu\nu}\partial_{(\mu}\bar{\Phi}\partial_{\nu)}\Phi -\mu^2 |\Phi|^2\left(1-\frac{2|\Phi|^2}{\sigma^2}\right)^2,
\label{eq:lagrangian}
\end{align}
where $g_{\mu\nu}$ is the metric with Ricci scalar $R$, and $\Phi$ is the complex scalar field making up the star with mass parameter $\mu$ and self-interaction strength $\sigma$. Here and in the following, we employ $G=c=1$ units. Boson stars are characterized by a harmonic dependence on both coordinate time $t$ and azimuthal coordinate $\varphi$, i.e., $\Phi\propto e^{i(\tilde{\omega} t+\tilde{m}\varphi)}$, set by their internal frequency $\tilde{\omega}$ and index $\tilde{m}$. Therefore, spinning stars (with $|\tilde{m}|>0$) exhibit toroidal surfaces of constant scalar field magnitude. So far, no binary boson star evolution resulted in remnants with stable light rings \cite{Cardoso:2016oxy,Palenzuela:2006wp,Palenzuela:2007dm,Palenzuela:2017kcg,Bezares:2017mzk,Helfer:2021brt,Bezares:2018qwa,Bezares:2022obu,Evstafyeva:2022bpr,Siemonsen:2023hko,Siemonsen:2023age,Sanchis-Gual:2018oui,Sanchis-Gual:2022mkk,Pierini:2022eim}. We proceed by focusing entirely on spinning stars \cite{Schunck1996,Volkov:2002aj,Kleihaus:2005me} with $\tilde{m}=3$ and $\sigma=0.035$. Stars in models with strong scalar self-interactions, e.g., $\sigma\ll 1$, as well as with larger $\tilde{m}$, exhibit larger compactnesses. The sequence of four binary boson stars considered in this work have mass-ratio $q=M/M_*$ and size-ratio $\eta=R/R_*$ between the primary (heavier) star of mass $M$ and radius $R$ and the secondary (lighter) binary constituent of mass $M_*$ and radius $R_*$. We define $M_0\approx M$ to be the ADM mass of the binaries for later convenience. In all cases, the primary exhibits large compactness $C=M/R$; the secondary is the same $C_*=0.02$ star in all cases. The properties of the binaries are summarized in \tablename{ \ref{tab:propMain}}. Isolated star solutions are used to generate constraint-satisfying binary initial data as described in Ref.~\cite{Siemonsen:2020hcg,Siemonsen:2023age}. All primary stars, and due to the large mass-ratio all merger remnants, exhibit a pair of counter rotating stable and unstable light rings. We focus solely on large mass-ratio systems, as the merger remnant of a comparable mass-ratio binary is likely a black hole (for further details see the Supplementary Material). In what follows, we identify the binaries by the compactness of the primary constituent. We also evolve a single black hole-boson star collision, where we replaced the $C=0.38$ primary star by a black hole of the same mass and spin. The initial coordinate separation is $D/M_0=160$ and the objects are boosted to Newtonian freefall velocities from infinity. Since the numerical evolutions are performed imposing axisymmetry on the metric and azimuthal symmetry $\partial_\varphi \Phi=3i\Phi$ on the scalar field, a non-axisymmetric instability \cite{Sanchis-Gual:2019ljs,Siemonsen:2020hcg} likely present in these stars is (artificially) quenched; see the Supplementary Material for further details.

\begin{figure}
\includegraphics[width=0.48\textwidth]{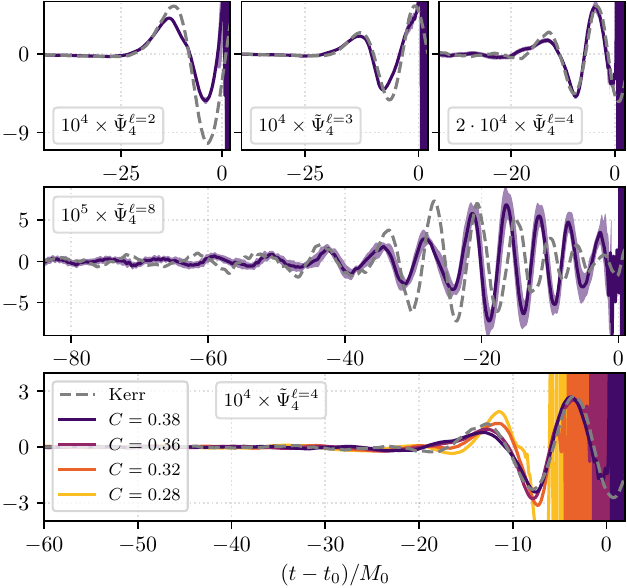}
\caption{The GWs emitted during the merger of the binaries labelled by their primary star compactness compared to the black hole-boson star binary labelled ``Kerr'' (top and center panels contain the $C=0.38$ and ``Kerr'' binaries only). We show the $\ell$th spin-weighted spherical harmonic component of the Newman-Penrose scalar $\tilde{\Psi}_4^\ell=r_{\rm extr.} M_0 \Psi_4^\ell$ extracted on spheres of coordinate radius $r_{\rm extr.}/M_0=100$ and with coordinate time $t$. All waveforms are time shifted by aligning at the largest peak of the $\ell=2$ mode, whereas $t_0$ corresponds to the transition time from prompt to subsequent emissions of the $C=0.38$ binary. The colored band indicates uncertainties of our methods.}
\label{fig:prompt}
\end{figure}

\begin{figure*}[t!]
\includegraphics[width=0.9\textwidth]{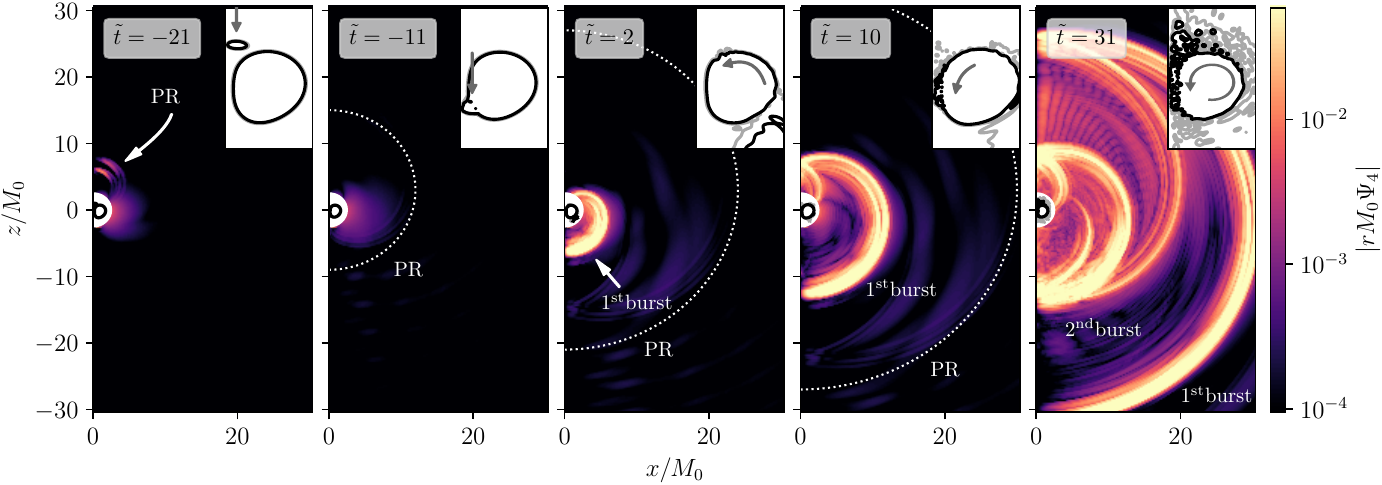}
\caption{The state of the $C=0.38$ binary system at selected coordinate times $\tilde{t}=(t-t_0+r_{\rm extr.})/M_0$ during the plunge, merger, and ringdown. The main plots show the Newman-Penrose scalar $\Psi_4$ at coordinate radii $r>3M_0$, which qualitatively shows the local GWs. We indicate the initial production and subsequent propagation of the high-frequency component of the prompt response as ``PR'' together with a dotted white circle (roughly corresponding to the propagating wavefronts). Similarly, the first two GW bursts are labelled $1^{\rm st}$ and $2^{\rm nd}$. In the region $r<3M_0$, we show two (black and gray) surfaces of constant scalar field magnitude $|\Phi|$ (with a close-up in the insets); recall, spinning boson stars are toroidally shaped. Arrows in the insets indicate the motion of the largest perturbation of $|\Phi|$ between snapshots. The symmetry axis is at $x=0$.}
\label{fig:dynamics}
\end{figure*}

\textit{Results} -- Our first main result is that the prompt dynamical response of the merger remnant encoded in the emitted GWs approaches the response of a Kerr black hole towards large-compactnesses. In \figurename{ \ref{fig:prompt}}, we show the waveforms emitted during the plunge and merger of the sequence of binaries and compare these with the aforementioned black hole-boson star binary. While there are differences in amplitude in the $\ell=2,3$ modes, the frequencies of all $\ell=2,3,4$ modes broadly match with the black hole response, and scale, analogous to black hole quasi-normal modes, (roughly) linearly with $\ell$. Since $C=0.38$ is still a relatively low remnant compactness, the peaks of the $\ell=2,3,4$ ``Kerr''-waveforms are burried in the subsequent emissions. However, we find higher-$\ell$ modes to peak earlier leading to a separation of the prompt response and subsequent emissions. Fitting for these $\ell=8$ frequencies and decay rates of the $C=0.38$ and ``Kerr'' cases (shown in \figurename{ \ref{fig:prompt}}), we find that the former differ by $\approx 15\%$, while the latter are consistent to within the uncertainties of our methods (though, the peak times differ by $\approx 5M_0$). We discuss these differences below. Lastly, in the bottom panel of \figurename{ \ref{fig:prompt}}, we present the $\ell=4$ waveforms for all four binaries. From there we conclude that the more compact the remnant (i.e., the longer the interior's light crossing time) the more and longer it's response is black hole-like. All in all, our nonlinear and self-consistent treatment of the large-mass-ratio problem broadly confirms expectations based on perturbative calculations.

To understand the system's dynamics during the plunge and merger, we present snapshots of the evolution of the $C=0.38$ case in \figurename{ \ref{fig:dynamics}}. Around $\tilde{t}\approx-25$ (the high-$\ell$ part of) the prompt Kerr response is produced as the secondary approaches the primary \footnote{Note, while the $C=0.38$ star exhibits no strictly bound polar null geodesics (i.e., a light sphere), there are ``quasi''-bound such geodesics as detailed in the Supplementary Material.} (first panel). While the latter propagates outwards, the binary merges and well-localized perturbations propagate along the symmetry axis through the remnant's interior (second panel). As these perturbations reach the opposing side and begin propagating poloidially around the remnant's outer edge, the first gravitational (and scalar) wave burst is emitted (third panel). Before the second burst is emitted, the perturbations propagate through the interior along the symmetry axis a second time as indicated by the gray arrows (fourth panel). The perturbations come around a third time in the last panel. At this stage, the perturbations dispersed into a collection of gravitationally bound states around the remnant (see the last panel of \figurename{ \ref{fig:dynamics}}). As a point of comparison, the light crossing time of null geodesics in the $C=0.38$ isolated solution traveling along the axis from $z=M/C$ to $z=-M/C$ as seen by distant observers is $\tau_{\rm axis}/M=19.6$.

\begin{figure*}[t!]
\includegraphics[width=0.7185\textwidth]{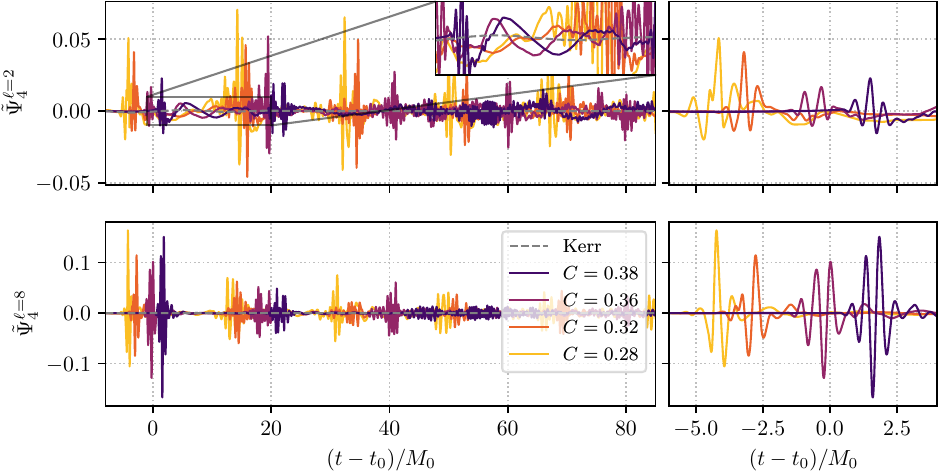}
\hfill
\includegraphics[width=0.2715\textwidth]{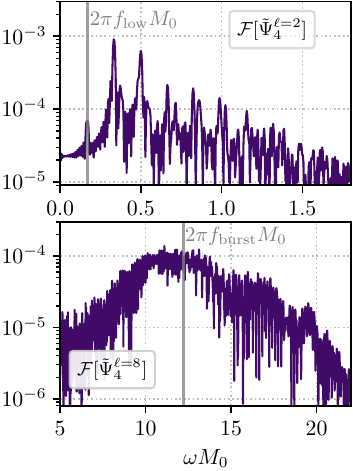}
\caption{ (left panels) The $\ell=2$ (top) and $\ell=8$ (bottom) components of the GWs emitted after the merger of the sequence of four binary boson stars compared with that of the black hole-boson star binary (labelled ``Kerr''); $t_0$ divides the prompt response and first burst of the $C=0.38$ binary, is defined as in \figurename{ \ref{fig:prompt}}. (central panels) Close-ups of the first bursts in each of the two polar modes. (right panels) The Fourier transform of the full $C=0.38$ signal as function of angular frequency $\omega$, focusing only on dominant radiation components in each mode: the long-lived component in the $\ell=2$ mode (top) and the burst-like high-frequency component in the $\ell=8$ mode (bottom); we indicate $f_{\rm low}$ and $f_{\rm burst}$. See the Supplementary Material for details on uncertainties of these waveforms.}
\label{fig:bursts}
\end{figure*}

\begin{figure}
\includegraphics[width=0.485\textwidth]{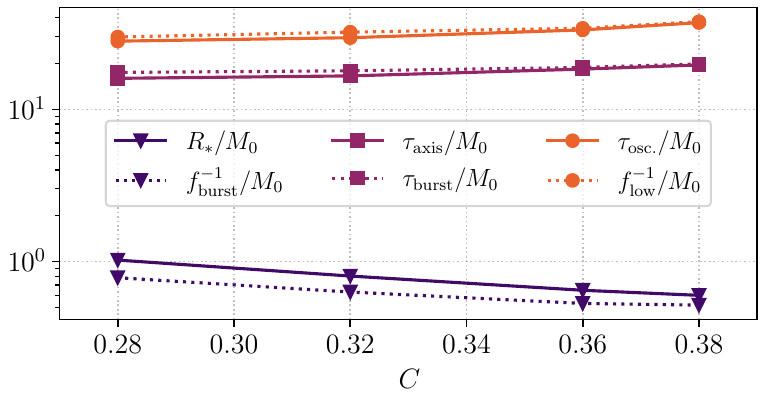}
\caption{The timescales associated with the gravitational waveforms emitted after the prompt emissions for all four binaries. We show the secondary's size $R_*\approx M_0/(\eta C)$, the frequency $f_{\rm burst}$ of the first burst, the light crossing time $\tau_{\rm axis}$, the burst period $\tau_{\rm burst}$, the dominant remnant oscillation period $\tau_{\rm osc.}$, and the frequency $f_{\rm low}$ of the long-lived component.}
\label{fig:timescales}
\end{figure}

Turning now to the gravitational waveform emitted after the prompt response. Our second main result is that these emissions exhibit bursts akin to GW echoes reappearing after roughly a light-crossing time through the remnants interior and with frequency set by the secondary's size. In \figurename{ \ref{fig:bursts}}, we present the $\ell=2$ and $\ell=8$ gravitational waveforms after the prompt emissions. Both modes contain a set of high-frequency bursts of frequency $f_{\rm burst}$ and separated by a timescale $\tau_{\rm burst}$. The burst's frequency can be understood as follows: The spatial scale $R_*\approx M_0/(\eta C)$ of the secondary object sets the lengthscale $\lambda$ of the perturbations sourced within the remnants interior: $\lambda\lesssim R_*$. In \figurename{ \ref{fig:timescales}}, we identify these perturbations as the source of the bursts, by finding good agreement between the underlying lengthscales, i.e., $f_{\rm burst}^{-1}\lesssim R_*$, for all four binaries considered. From the central panels of \figurename{ \ref{fig:bursts}}, we see that in contrast to the black hole ringdown frequencies, the burst frequency is independent of the polar mode number $\ell$. The burst period $\tau_{\rm burst}$ is less straightfoward to understand. At the test-field level, high-frequency gravitational perturbations sourced by the merger are propagating roughly along null geodesics. Therefore, most naturally the burst period is compared to the light crossing time of null geodesics traversing the remnants interior along the direction of the perturbations sourced during the merger, i.e., $\tau_{\rm axis}$, as defined above. In \figurename{ \ref{fig:timescales}}, we compare $\tau_{\rm burst}$ against this light crossing time $\tau_{\rm axis}$, again finding good agreement $\tau_{\rm burst}\approx \tau_{\rm axis}$. However, $\tau_{\rm axis}$ is to be understood only as a rough measure of crossing times, since purely considering null geodesics to determine the burst period neglects that during the merger massless (gravitational) and massive (scalar) modes are coupled and the relevant dynamics are not confined only to the axis. Overall, the appearence of burst-like emission components separated by a light crossing time of the remnants interior is consistent with the expectations based on test-field calculations. On the other hand, the burst's frequency $f_{\rm burst}\gtrsim \mathcal{O}(1)M_0> f_{\ell m=20}^{\rm Kerr}$, at the considered size-ratios, is much larger than the corresponding black hole quasi-normal mode frequencies \cite{Berti:2009kk,Berti:2005ys}, and hence, larger than test-field computations predict the GW echo frequency, and decay rates, to be \cite{Mark:2017dnq,Wang:2018gin,Maggio:2020jml}; this is further discussed below.

Our third main result is that the low-multipole GW emissions contain a large-amplitude and long-lived component. As evident from \figurename{ \ref{fig:bursts}} (zoom-in of the top left panel and Fourier transform in the top right panel), the $\ell=2$ mode exhibits low-frequency oscillations between the high-frequency burst. Physically, this quadrupolar contribution originates from oscillation modes of the remnant star excited during the merger. In \figurename{ \ref{fig:timescales}}, we compare the fundamental frequency $f_{\rm low}$ of this component to the oscillation period $\tau_{\rm osc.}$ of the remnants (measured from post-merger oscillations of $\max|\Phi|$), finding good agreement as a function of remnant compactness: $\tau_{\rm osc.}\approx f^{-1}_{\rm low.}$. The signal, $\tilde{\Psi}_4$, is dominated by the $n=1$ harmonic of $f_{\rm low}$, where each harmonic has frequency $(n+1)f_{\rm low}$ (see also the Fourier transform in \figurename{ \ref{fig:bursts}}). Notice also that $f_{\rm low}\approx f_{\ell m=20}^{\rm Kerr}/2$. Roughly, the dominant frequency in the $\ell$th mode of this second GW component scales as $f^{\ell>2}_{\rm low}\sim \ell$. As can be seen in \figurename{ \ref{fig:bursts}}, this component is long-lived, with longer decay timescales than our simulations. The equally spaced frequency spectrum, the long-lived nature, together with the observation $f^{-1}_{\rm low}\approx 2\tau_{\rm burst}$, are indicative of trapped modes, which objects with stable light rings generally exhibit \cite{Kokkotas:1999bd} (see also Refs.~\cite{Macedo:2013jja,Wang:2019rcf,Heidmann:2023ojf}). Lastly, we find no evidence that the amplitude of this long-lived component decreases with increasing compactness.

The amplitude of $\tilde{\Psi}_4$, however, should be interpreted with care. The observationally relevant strain $h$ scales as $h(\omega)\sim \tilde{\Psi}_4(\omega)/\omega^2$ in the Fourier domain. Hence, high-frequency features are suppressed by a factor of $\sim \omega^{-2}$ compared to low-frequency components of $\tilde{\Psi}_4$. Therefore, and this must be emphasized, at the level of the strain the amplitude of the long-lived component is \textit{larger} than the amplitude of the bursts for all $\ell\lesssim 6$ modes (and larger than the prompt response in all considered modes). In turn, the amplitude of the bursts is larger than the prompt response for $\ell\gtrsim 4$ only.

Lastly, the total GW energy $E_<(t)$ radiated through a sphere of radius of $r_{\rm extr.}$ up to time $t$ for the $C=0.38$ binary is $E_<(t_0+\tau_{\rm burst})/M_0\approx 6\times 10^{-6}$ and $E_<(t_0+5\tau_{\rm burst})/M_0=6\times 10^{-5}$. The energy $E_<(t)$ still increases roughly linearly with rate $\dot{E}_<\approx 8\times 10^{-7}$ at this time. In contrast, the total radiated energy of the head-on collision of a (non-spinning) binary black hole with the same mass-ratio is $\lim_{t\rightarrow \infty} E_<(t)= 2\times 10^{-6} M_0$ \cite{Sperhake:2011ik}; this is consistent with the energy emission by the black hole-boson star merger considered in this work.

\textit{Discussion \& Conclusion} -- In this work, we performed the first nonlinear and self-consistent study of the ringdown of a black hole mimicker. We broadly confirm the appearence of features in the emitted gravitational waveform expected based on test-field approaches. However, we also found that the dynamics of the mimicker's internal degrees of freedom have large impact on the emitted GWs. 

Although, the sequence of remnants, considered in this work, becomes increasingly more compact, there are no polar light spheres, and the primary's exterior is (by construction) not parametrically close to the Kerr exterior. Therefore, it is remarkable that the waveforms in \figurename{ \ref{fig:prompt}} match at all, and in principle, any difference can be attributed to the differing exteriors. In analogy to the black hole case \cite{Davis:1971gg,Sperhake:2011ik,East:2013iwa,East:2014nfa}, the small compactness of the secondary, $C_*=0.02$, likely has small impact on the prompt response of the remnant in low-$\ell$ modes.

GW echoes---features of the scattering of massless waves in black hole mimicker spacetimes neglecting any kind of backreaction---are expected to exhibit frequencies and decay rates of the order of the Kerr quasi-normal modes at early times~\cite{Mark:2017dnq,Vellucci:2022hpl,Maggio:2019zyv}. In this work, we identified \textit{two} distinct frequency components of the waveform emitted after the prompt response. One of these is consistent with the quasi-normal mode frequencies of a black hole of the same mass and spin, but long-lived, while the other is burst-like in nature, but with frequencies much higher than those quasi-normal frequencies. Therefore, either component could \textit{ambiguously} be labelled ``the echoes''. Since the bursts originate in perturbations of the mimicker's internal degrees of freedom and scale with the size of the secondary, test-field approaches are generally insufficient to capture this component \footnote{In fact, naively applying these perturbative methods, one would \textit{not} expect the waveforms to exhibit echoes, because the boson star's light-crossing time is comparable to Kerr quasi-normal mode decay timescales.}. Lastly, we find no evidence of a nonlinear instability in the binary constituents or remnants \cite{Keir:2014oka,Cardoso:2019rvt,Cunha:2022gde}.


\begin{acknowledgments}
\textit{Acknowledgments} -- We would like to thank Niayesh Afshordi, Alejandro C\'{a}rdenas-Avenda\~{n}o, Will East, Suvendu Giri, Luis Lehner, Elisa Maggio, and Frans Pretorius for many interesting discussions about aspects of this work. We especially thank Will East and Frans Pretorius for comments on an earlier version of this draft. The author is pleased to acknowledge that the work reported on in this paper was substantially performed using the Princeton Research Computing resources at Princeton University which is a consortium of groups led by the Princeton In- stitute for Computational Science and Engineering (PICSciE) and Office of Information Technology’s Research Computing. This work used \texttt{anvil} at Purdue University through allocation PHY230198 from the Advanced Cyberinfrastructure Coordination Ecosystem: Services \& Support (ACCESS) program \cite{access}, which is supported by National Science Foundation grants \#2138259, \#2138286, \#2138307, \#2137603, and \#2138296.
\end{acknowledgments}

\nocite{*}

\bibliography{bib.bib}

\appendix
\section{Supplemental Material}

\section{Boson stars} \label{app:bs}

\subsection{Parameter space}

\begin{figure}[t]
\includegraphics[width=0.49\textwidth]{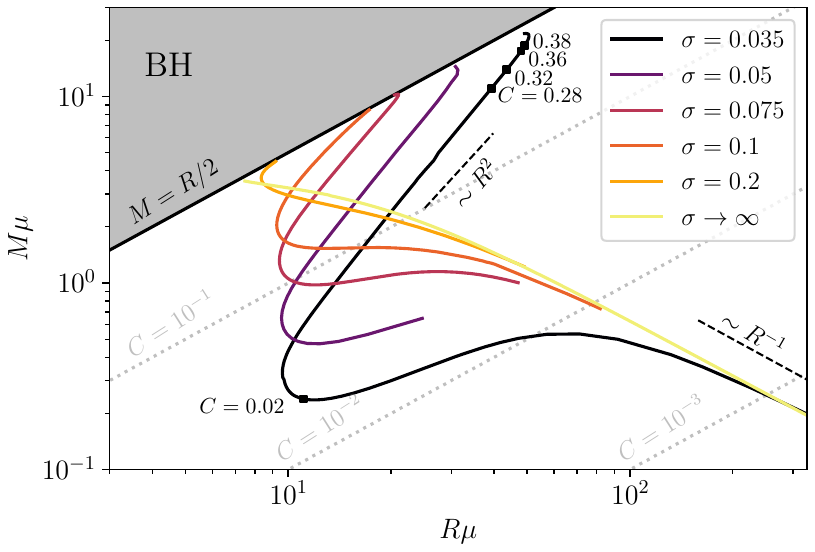}
\caption{We show the masses and radii (measured with respect to the scalar mass parameter $\mu$) of several $\tilde{m}=3$ families of rotating boson stars in the scalar model with self-interaction strength $\sigma$ (see, eq.~\eqref{eq:lagrangian}). In the decoupling limit $\sigma\rightarrow\infty$, the scalar potential reduces to the mass term $\mu^2|\Phi|^2$. Regions of the parameter space with compactness $C=M/R>1/2$ are labelled with ``BH''. Two relevant radius scalings are shown by dashed black lines in the respective regimes. Lastly, we indicate the locations of those rotating stars considered in this work by black squares and label each with their compactness.}
\label{fig:Comp}
\end{figure}

The rotating boson stars relevant for this work are asymptotically flat horizonless regular axisymmetric and stationary solutions to the Einstein-Klein-Gordon equations associated with \cite{Schunck1996,Volkov:2002aj,Kleihaus:2005me}
\begin{align}
\mathcal{L}=\frac{R}{16\pi}& -g^{\mu\nu}\partial_{(\mu}\bar{\Phi}\partial_{\nu)}\Phi -\mu^2 |\Phi|^2\left(1-\frac{2|\Phi|^2}{\sigma^2}\right)^2,
\label{eq:lagrangian}
\end{align}
where $g_{\mu\nu}$ is the metric with Ricci curvature scalar $R$, and $\Phi$ is the complex scalar field making up the star with mass parameter $\mu$ and self-interaction strength $\sigma$. As noted in the main text, the full scalar field profile making up an isolated star takes the form $\Phi=\phi(r,\theta)e^{i(\tilde{\omega} t+\tilde{m}\varphi)}$, with azimuthal index $\tilde{m}$ and internal frequency $\tilde{\omega}$. The metric ansatz for rotating solutions, $|\tilde{m}|\geq 1$, in Lewis-Papapetrou form is given by
\begin{align}
\begin{aligned}
ds^2=-fdt^2 & +\frac{l}{f}\bigg[g(dr^2 +r^2d\theta^2)\\
& +r^2\sin^2\theta\Big(d\varphi-\frac{\Omega}{r}dt\Big)^2\bigg],
\end{aligned}
\label{eq:metricansatz}
\end{align}
where $f,l,g$, and $\Omega$ depend on $r$ and $\theta$ only. Regularity on the symmetry axis implies that the scalar field magnitude of rotating stars vanishes there. At large distances, the magntiude $\phi$ decays exponentially $\sim e^{-br}$, for some $b>0$. The mass $M$ and angular moment $J$ of these stationary and axisymmetric spacetimes are given by their corresponding Komar expressions. We define the radius of a given solution to be the circular coordinate radius $R$ at which $99\%$ of the mass of the solutions lies at $r<R$ \cite{Siemonsen:2020hcg}. 

In \figurename{ \ref{fig:Comp}}, we show the masses and radii of $\tilde{m}=3$ families of spinning boson star solutions in scalar models with varying coupling $\sigma$. In all cases, the scalar self-interactions are irrelevant in the Newtonian limit, where solution's masses scale as $M\sim R^{-1}$ and the compactness is small. Towards the relativistic regime, the properties of families in models with different coupling strength $\sigma$ diverge. Importantly, if self-interactions are neglected, i.e., $\sigma\rightarrow\infty$, then the solution's mass scales roughly as in the Newtonian limit, even in the relativistic regime. In contrast, the energy density of those rotating solutions in strongly coupled scalar models is dominated by surface tensions in this thin-wall limit \cite{Lee:1986ts,Friedberg:1986tq,Lee:1991ax}, scaling as $M\sim R^2$. Understanding these scalings is important for the choice of binary boson star to model the ringdown of a black hole mimicker in the large-mass-ratio regime, as we point out in the next section. Families of solutions with higher $\tilde{m}$ are more rapidly spinning, supporting more compact stars. This is the primary reason for considering $\tilde{m}=3$, rather than lower-$\tilde{m}$ solutions. One drawback of this choice is that $|\tilde{m}|>1$ solutions are likely linearly unstable to a non-axisymmetric instability \cite{Sanchis-Gual:2019ljs} even in the small-$\sigma$ regime \cite{Siemonsen:2020hcg}. In our evolution setup (enforcing the metric's axisymmetry and the scalar azimuthal symmetry), however, this instability is removed. Therefore, as there are no ergoregion present in the considered boson stars, all star solutions relevant in this work are stable against any known linear instabilities (in our evolution setup). 

\subsection{Choice of binary boson stars}

We are interested in studying the merger and ringdown of black hole mimicking objects using boson stars. Utilizing these solutions, however, to model mimickers comes with two major limitations: (i) as these are entirely classical solutions, the merger remnant of two of these objects collapses to a black hole roughly when its mass surpasses the hoop bound \cite{1973grav.book.....M} or is above the maximum mass of the family of boson star solutions, and (ii), within a given scalar theory (at fixed $\mu$ and $\sigma$), sequences of solutions follow the scaling $M\sim R^p$ with $p\neq 1$ (i.e., not the black hole scaling $M\sim R$). The former implies that one likely cannot consider equal-mass mergers of ultra compact boson stars, since their merger products are always black holes or less compact stars (see e.g., Ref.~\cite{Cardoso:2016oxy} or Fig. 6 in Ref.~\cite{Siemonsen:2023age}). Hence, in order to circumvent this issue, we consider the large-mass-ratio regime, in which a primary ultra compact star is perturbed by a lighter secondary star. We find that a star of compactness $C\approx 0.39$, in the same sequence of solutions indicated in \figurename{ \ref{fig:Comp}}, collapses to a black hole shortly after merger with the $C_*=0.02$ secondary considered throughout this work.

Issue (ii) implies that, at fixed mass-ratio, only one of the binary's constituents may be ultra compact. In this work, we are interested in constructing binaries, which model the merger of two ultra compact objects in the large-mass-ratio regime the ``closest''. Hence, ideally one would like to use two objects of large compactness each; this, however, is not achievable in the context of boson stars due to the restrictions shown in \figurename{ \ref{fig:Comp}} (i.e., nowhere does the boson star mass scale as $\sim R$). Let us consider binaries built from stars with the Newtonian scaling, $M\sim R^{-1}$, in the large-mass-ratio regime $M_1/M_2\gg 1$. Then the size of the \textit{lighter} companion star would be \textit{larger} than the primary highly-compact star, since $R_1/R_2\sim M_2/M_1\ll 1$. In this regime, even if the primary were a black hole, the emitted gravitational radiation is not expected to exhibit a black hole ringdown \cite{East:2014nfa}. Therefore, we focus on those solutions, which are in the thin-wall regime of the parameter space instead. As pointed out above, there the masses along a family of solutions scale as $M\sim R^2$, which is much closer to the black hole $M\sim R$ scaling behavior. Ultimately, a scalar potential yielding boson star solutions with $M\sim R$ in the relativistic regime would be most relevant (see Ref.~\cite{Pitz:2023ejc} for a systematic study in the non-spinning case).

\subsection{Null geodesics}

\subsubsection{Equatorial light rings}

In order to understand the propagation of null geodesics through the boson star spacetimes considered in this work, we begin by analyzing the relevant potentials for radial motion of the latter confined to the equatorial plane. 
To that end, let $\lambda$ be an affine parameter of the trajectory $x^\mu(\lambda)$ of the null geodesic with tangent $\dot{x}^\mu(\lambda)=dx^\mu/d\lambda=u^\mu(\lambda)$, such that $u_\mu u^\mu=0$. The conserved energy, $E=-t_\mu u^\mu$, and angular momentum, $L_z=\varphi_\mu u^\mu$, of the geodesic are measured with respect to the Killing fields associated with stationarity and axisymmetry of the isolated boson star spacetimes, $t^\mu$ and $\varphi^\mu$, respectively. The normalization condition $u_\mu u^\mu=0$ yields
\begin{align}
\dot{r}^2=-\frac{f}{gl}V, & & V = -\frac{\kappa L_z^2}{f}(\kappa-H_+)(\kappa-H_-),
\end{align}
where the impact parameter is $\kappa=E/L_z$, and the potentials $H_\pm$ are \cite{Cunha:2016bjh,Cunha:2017qtt}
\begin{align}
H_\pm=\frac{\Omega}{r}\pm\frac{f}{r\sin\theta\sqrt{l}}, 
\label{eq:hphm}
\end{align}
in the Lewis-Papapetrou coordinates chosen above. Light rings inside the equatorial plane ($\theta=\pi/2$ and $\dot{\theta}=0$) are defined by $V=0$ and $\partial_r V=0$ (the latter is equivalent to $\partial_rH_\pm=0$). Furthermore, roots of $\partial_rH_+$ ($\partial_rH_-$) correspond to co-(counter-)rotating light rings, and these are stable if $\pm\partial_r^2H_\pm>0$ at the location of the light ring (or unstable if $\pm\partial_r^2H_\pm<0$).

\begin{figure}
\includegraphics[width=0.485\textwidth]{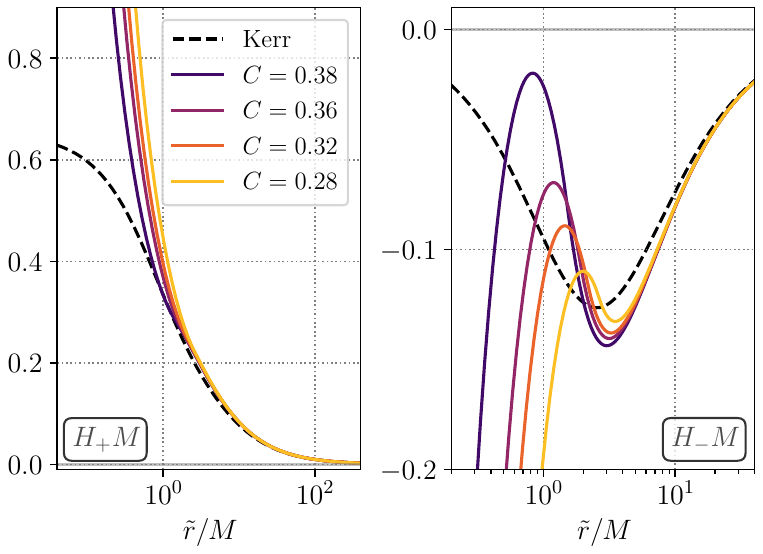}
\caption{The potentials $H_\pm$, defined in \eqref{eq:hphm}, characterizing the structure of light rings in the equatorial plane of a Kerr black hole with spin $a/M=0.95$ (labelled ``Kerr''), and the sequence of spinning boson stars of increasing compactness considered throughout this work (labelled by their compactness). The radial coordiante $\tilde{r}$ is the coordinate $r$ in the case of boson stars and $r-r_H$ in the case of Kerr [in the coordinates of \eqref{eq:metricansatz}], where $r_H$ is the radial location of the horizon.}
\label{fig:hphm}
\end{figure}

In \figurename{ \ref{fig:hphm}}, we compare the potentials $H_\pm$ of a Kerr black hole with spin $a/M=0.95$ with those of the sequence of stars of increasing compactness indicated in \figurename{ \ref{fig:Comp}}. There exists a co- and counter rotating (unstable equatorial) light ring around Kerr black holes. In this large-spin regime, these light rings are located at $r/M\approx 1$ and $r/M\approx 4$ (corresponding to $\tilde{r}/M\approx 0$ and $3$ in \figurename{ \ref{fig:hphm}}), respectively. Turning now to the sequence of spinnning boson stars, all solutions in the sequence exhibit a stable and unstable counter-rotating light ring in the equatorial plane (and no co-rotating light ring). The orbital frequencies $d\varphi/dt=\tilde{\Omega}$ of the stable light rings in the boson star spacetimes are $M\tilde{\Omega}_S \times 10^2=2,5,9,11$, as well as $M\tilde{\Omega}_U=0.1437,0.142,0.138,0.132$ for the unstable light rings, in decreasing star compactness $C=0.38,0.36,0.32,0.28$, respectively. Interestingly, the orbital frequency of the unstable light ring in a Kerr spacetime of spin $a/M=0.95$ is $M \tilde{\Omega}_{\rm Kerr}=0.1446$. Therefore, the relative difference of the unstable light ring's frequency between the $C=0.38$ rotating boson star and the Kerr spacetime with the same angular momentum is $<1\%$. The absence of a co-rotating light ring may be understood intuitively as follows: Along the considered sequence of rotating boson stars, and with increasing compactness, the outer light ring (i.e., the counter-rotating ring) appears first, whereas the inner, co-rotating, light ring may appear only for more compact solutions. 

\subsubsection{Polar null orbits}

In the case of ultra compact and spherically symmetric boson stars \cite{Palenzuela:2017kcg,Boskovic:2021nfs,Collodel:2022jly} the presence of equatorial light rings implies the existence of complete photon spheres. As rotating boson stars are not connected to their non-rotating counterparts by continuously increasing the angular momentum \cite{Kobayashi:1994qi,Kleihaus:2005me}, the existence of polar null orbits, i.e., those bound null geodesics with $L_z=0$, is not implied by the presence of equatorial light rings. In the following, we briefly investigate the existence and properties of the bound polar null geodesics in the sequence of rotating boson stars (and beyond) indicated in \figurename{ \ref{fig:Comp}}. 

\begin{figure}
\includegraphics[width=0.485\textwidth]{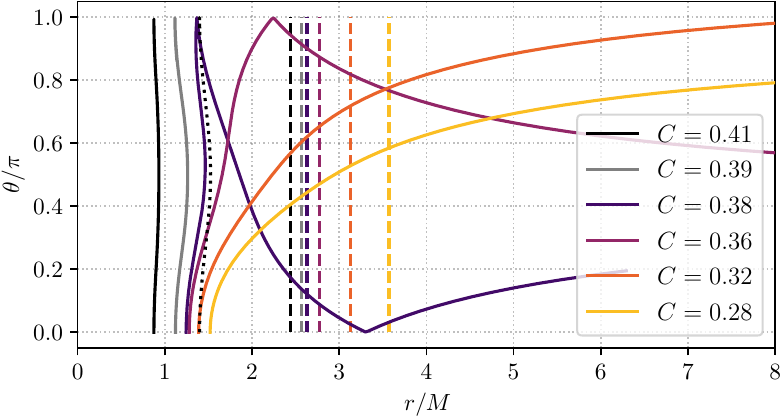}
\caption{The polar and radial coordinates $\theta$ and $r$, respectively, of the polar null geodesic, which maximizes $\tau_{\rm polar}$. The boson stars are identified by their compactness $C$. Those with $C\leq 0.38$ are listed in \tablename{ \ref{tab:propMain}}, and have been evolved in the context of binaries in the bulk of this work. Those rotating stars with $C\geq 0.39$ are slightly more compact representatives of the same family of solutions, which have frequencies $\tilde{\omega}/\mu=0.101$ and $\tilde{\omega}/\mu=0.095$. Furthermore, dashed lines indicate the coordinate radius of each star $r/M=1/C$ and the black dotted (solid) line, corresponds to the outer (inner) closed null polar geodesic of the $C=0.41$ spacetime.}
\label{fig:rvstheta}
\end{figure}

Due to the lack of a Killing tensor, and associated constant of motion, we proceed by numerically solving the geodesic equation to find closed polar null geodesics (with further details on the numerical implementation deferred to Appendix~\ref{app:num}). We integrate starting on the spin axis (with vanishing initial radial velocity) of each of the stars indicated in \figurename{ \ref{fig:Comp}}. This is performed for various initial positions along the symmetry axis. We define $\tau_{\rm polar}$ as the coordinate time such a geodesic spends inside the star, $r<M/C$. More generally, and for later convenience, this coordinate time is defined as
\begin{align}
\tau=\int_{\gamma} \frac{dt}{d\lambda}d\lambda.
\label{eq:tau}
\end{align}
Then for those initial positions along the symmetry axis approaching a bound null geodesic the time $\tau_{\rm polar}$ diverges. If a bound orbit exists, then iteratively increasing this time results in the convergence of the initial position along the symmetry axis towards that orbit. If no bound polar null geodesics exist, this proceedure yields the geodesic with largest $\tau_{\rm polar}$.

In \figurename{ \ref{fig:rvstheta}}, we show those polar null geodesics, which maximize $\tau_{\rm polar}$. The time $\tau_{\rm polar}$ increases along the sequence of boson stars with increasing compactness. In the case of the $C=0.38$ star, the polar null geodesic maximizing $\tau_{\rm polar}$ passes through the spin axis twice before escaping to infinity. The sequence of stars indicated in \figurename{ \ref{fig:Comp}} are elements of a family of solutions with more compact stars (see also \figurename{ \ref{fig:Comp}}). Continuing further along this family of solutions, the $C=0.39$ boson star exhibits at least one polar light ring\footnote{With our numerical methods we can confidently identify only the existence of a single polar orbit. It is plausible that two distinct orbits exit, which we are unable to distinguish due to their relative proximity.}, whereas the one with $C=0.41$ has two distinct bound polar null orbits. This demonstrates that the family of solutions indicated in \figurename{ \ref{fig:Comp}} develops bound polar null orbits in the high-compactness limit. On the other hand, those rotating boson stars evolved within binaries in this work, exhibit only ``quasi''-bound null orbits, with large, but finite, $\tau_{\rm polar}$. Explicitly, the coordinate times are $\tau_{\rm polar}/M=28.8,17.5,11.7,9.8$ for the stars of compactness $C=0.38,0.36,0.32,0.28$, respectively.

\subsubsection{Light-crossing times}

The high-frequency perturbations sourced during the merger of the binary propagate through the remnant star's interior along the symmetry axis, partially reflect on the opposite side and propagate backwards both along the symmetry axis and along the surface of the remnant. In the eikonal limit, massless perturbations propagate along null geodesics through the spacetime \cite{1972ApJ...172L..95G,Isaacson:1968hbi}. Within our numerical setup, we enforce axisymmetry of the metric and a $\tilde{m}=3$ azimuthal symmetry for the scalar field \footnote{Recall, this corresponds to $\mathcal{L}_k\Phi=i\tilde{m}\Phi$ with $\tilde{m}=3$ and the Lie derivative $\mathcal{L}_k$ along the Killing vector $k^\mu$ generating the spacetime's axisymmetry.}. Therefore, geodesics with vanishing angular momentum, $L_z=0$, are particularly important to our discussion, as these correspond to the propagation of high-frequency and $m=0$ massless modes (the equatorial light rings correspond to modes with $m\gg 1$). 

Accurately describing the wave dynamics in a merger setting is non-trivial. Therefore, in the following we determine the light crossing times of the interior of the sequence of rotating boson stars indicated in \figurename{ \ref{fig:Comp}} along $L_z=0$ null geodesics simply as a rough measure of the time delay of the gravitational wave bursts emerging during the ringdown. To that end, we focus on three classes of such geodesics: (I) ``quasi''-bound polar null geodesics (as defined above), (II) equatorial null geodesics propagating from the unstable light ring through the spin axis to the opposite side, and (III) massless mode propagation along the symmetry axis. In all cases, we use \eqref{eq:tau} to determine the light crossing times with respect to observers at infinity. 

\begin{figure}
\includegraphics[width=0.485\textwidth]{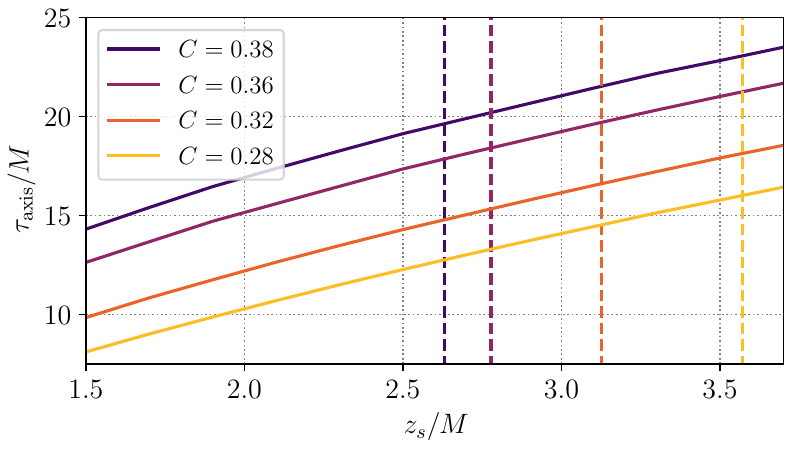}
\caption{The light crossing time $\tau_{\rm axis}$ as a function of start/endpoint $|z_s|$ along the symmetry axis for all stationary ultracompact solutions considered in this work. The coordinate radii $z_s/M=1/C$ are indicated by vertical dashed lines.}
\label{fig:tau_lc}
\end{figure}

The crossing times of geodesics of type I were determined in the previous section to be $\tau_{\rm polar}\sim \mathcal{O}(10)M$. The crossing time of type-II or those of geodesics within the equatorial plane emerging from the unstable null orbit (as found in \figurename{ \ref{fig:hphm}}), passing through the symmetry axis, and ending on the unstable null orbit (on the opposite side of the object). These times are $\tau_{\rm equa.}/M=12.2,11.8,10.1$, and $9.2$, corresponding to the stars of compactness $C=0.38,0.36,0.32$, and $0.28$, respectively. Note, the light-crossing time through the interior increases with decreasing radius $R=M/C$. Lastly, the light crossing time $\tau_{\rm axis}$ associated with geodesics of category III are less naturally bounded. Therefore, we determine $\tau_{\rm axis}$ for geodesics along the spin axis starting at various coordinates $z_s>0$ above the equatorial plane and terminating at $-z_s$. In \figurename{ \ref{fig:tau_lc}}, we show the dependence of $\tau_{\rm axis}$ on this choice of $z_s$.

\section{Numerical implementation \& uncertainties} \label{app:num}

\subsection{Numerical evolution methods \& convergence}

The isolated rotating boson stars making up the binaries were obtained using methods developed in Refs.~\cite{Siemonsen:2020hcg}. Specifically, we use Newton-Raphson relaxation techniques in conjunction with fifth-order finite difference methods to solve the set of two-dimensional elliptic partial differential equations emerging, when plugging the scalar field ansatz and \eqref{eq:metricansatz} into the Einstein-Klein-Gordon equations derived from \eqref{eq:lagrangian}. The equations are discretized uniformly in the polar coordinate $\theta$ and compactified radial coordinate $\bar{r}=r/(1+r)\in (0,1)$. Obtaining boson star solutions in the thin-wall regime is challenging due to large spatial gradients developing at the star's surface. We overcome some of these issues by rescaling the radial coordinate $\bar{r}$ with lengthscale $L$, i.e., $\bar{r}\rightarrow L \bar{r}$, as well as all other dimensionful quantities, in order for the star's surface to reside (roughly) at $\bar{r}\approx 0.5$. This ensures that the gradients at the star's surface are best-resolved at fixed resolution. The resolutions used for all boson stars in the radial and polar directions are $N_{\bar{r}}\times N_\theta=1500\times 100$. For all highly compact binary constituents indicated in \figurename{ \ref{fig:Comp}}, we use $L=0.025$. Note, in the case of the $C_*=0.02$ secondary companion, $L=1$ is entirely sufficient. 

These isolated solutions are combined into binaries using techniques developed in Refs.~\cite{Siemonsen:2023age,East:2012zn}. In particular, we solve for constraint satisfying binary boson star initial data using the conformal thin-sandwich formulation of the Hamiltonian and momentum constraints of the Einstein equations. The necessary free data is obtained by superposing two displaced and boosted star solutions. The scalar kinetic energy of the binary (entering the constraint equations) is conformally rescaled with power $p=-3$, as defined in detail in Ref.~\cite{Siemonsen:2023age}, to reduce spurious oscillations of the stars in the subsequent evolution of the initial data. Note, for the black hole-boson star binary, we use plain superposed initial data without solving the constraints explicitly. In this latter case, we verified that the total charge $Q$ [associated with the global U(1) symmetry of \eqref{eq:lagrangian}] of the initial data agrees with the total charge of the isolated secondary to within $<1\%$ relative difference (and is stable throughout the entire evolution until merger). The black hole's properties differ from those parameters initialized in the initial data only by $<0.1\%$ throughout the evolution up until the merger.

We proceed to evolve these initial data with the same methods as used in Refs.~\cite{Siemonsen:2020hcg,Siemonsen:2023hko}. We numerically solve the full Einstein-Klein-Gordon equations of motion, derived from \eqref{eq:lagrangian}, using the generalized harmonic formulation \cite{Pretorius:2004jg}. The equations are discretized using fourth-order accurate finite differences in conjunction with fourth-order Runge-Kutta time stepping. All evolutions are performed assuming axisymmetry for the metric, and azimuthal symmetry for the scalar field (as defined above), which is achieved numerically by means of a generalized Cartoon method \cite{Pretorius:2004jg,Alcubierre:1999ab}. Essential is the role of the adaptive mesh refinement (AMR) with refinement ratio 2:1 \cite{East:2011aa}. Due to the large separation of scales of the problem, we employ nine mesh refinement levels with a medium spatial resolution of $\Delta x_{\rm finest}/M_0\approx 3\times 10^{-3}$ on the finest level (covering both the primary and secondary of each binary). Crucially, due to the high-frequency nature of the emitted burst-like gravitational waves after the merger of the binary, we ensure the wave extraction zone (out to radial coordinate distances $r/M_0\approx 133$) is well-resolved with grid spacing of $\Delta x_{\rm wavezone}/M_0\leq 0.025$ at medium resolution. For all evolutions, we utilize constraint damping terms with damping rate $\kappa_D M_0=3$ and damping constant $\rho_D=0.75$ \cite{Gundlach:2005eh}. For all boson star binaries, we use stationary gauge throughout the evolution to minimize spurious gauge dynamics. For the black hole-boson star we employ the damped harmonic gauge condition \cite{Choptuik:2009ww,Lindblom:2009tu}. 

\begin{figure}
\includegraphics[width=0.48\textwidth]{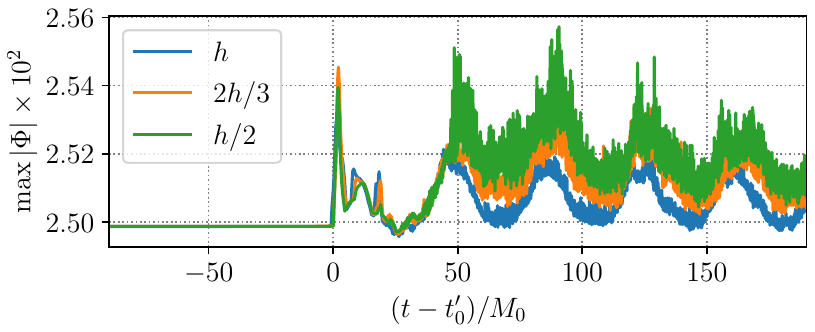}
\includegraphics[width=0.48\textwidth]{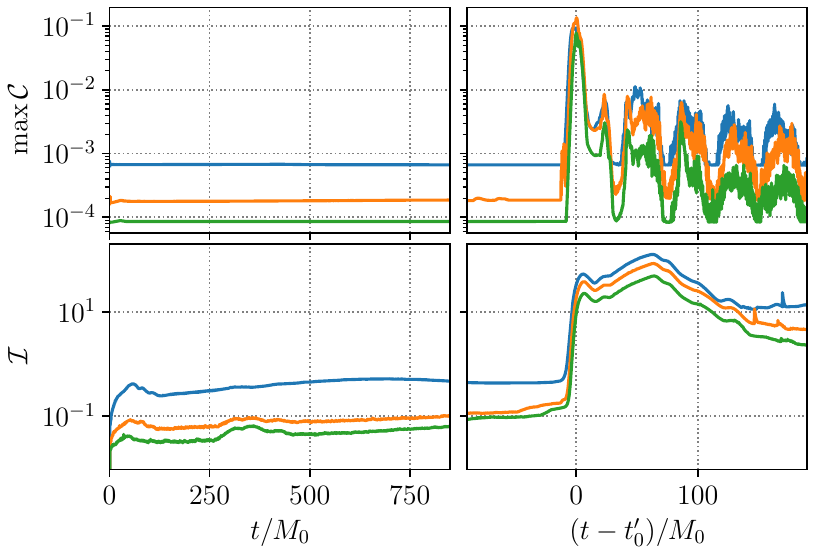}
\caption{The convergence behavior of constraint violations, measured by $\max\mathcal{C}$ and $\mathcal{I}$ (defined in the main text), throughout the evolution of the binary $C=0.38$ with decreasing grid spacing from $h$ at low resolutions to $h/2$ at high resolution. (top) The global maximum of the scalar field magnitude throughout the plunge and merger of the binary (all curves are aligned at the time of merger $t'_0$). The evolution of the global maximum $\max\mathcal{C}$ (center row) and the integrated norm $\mathcal{I}$ (bottom row) of the constraint violations during the initial phase before the plunge (left column), as well as during and after the merger (right column). Note, the merger occurs at $t/M_0\approx 10^3$.}
\label{fig:conv}
\end{figure}

To validate our findings, we perform a convergence study on the boson star binary with the most compact constituent (i.e., the $C=0.38$ binary), which exhibits the largest separation of scales and most extreme remnant. To that end, we use the supremiums norm $\max \mathcal{C}$ and integrated norm $\mathcal{I}=\int d^3x\sqrt{\gamma}\mathcal{C}/M_0^3$ (integrated over a ball of coordinate radius $64M_0$ centered around the center of mass), with $\mathcal{C}=\sum_\mu|H_\mu-\square x_\mu |M_0/4$, to track the constraint violations throughout the evolution. In \figurename{ \ref{fig:conv}}, these quantities are shown as a function of time at low, medium (default) and high resolutions. Early on in the evolution, the constraint violations converge towards zero at convergence order between third and the expected fourth order. Note, the convergence of $\max\mathcal{C}$ slows towards higher resolutions. This is primarily due to the resolution limitations to solve for the isolated solution of the ultra compact constituent as pointed out above. At merger, $t\approx t'_0$, short-wavelength modes are excited in the remnant resulting in worse convergence behavior (i.e., the wavelength of some of these modes approaches the grid scale) in both $\mathcal{I}$ and $\max\mathcal{C}$. At this time, the integrated norm converges roughly at first order between the low and medium resolution runs (zeroth order for the supremums norm), and at second order between medium and high resolution cases (first order for the supremums norm). This suggests that, likely driven by high-frequency modes, the low resolution is not yet in the converging regime. We use the medium resolution as the default resolution for all binaries considered in this work. In the following section, we discuss the impact of resolution on the extract gravitational waveforms.

\subsection{Gravitational waveform uncertainties}

To understand the numerical uncertainties associated with observables presented in the bulk of this work, we estimate the impact of different sources of error on the extracted gravitational waveform. In particular, due to the largest mass- and size-ratio the $C=0.38$ binary configuration is the most challenging to treat within our numerical setup. Hence, we focus on the uncertainties of this binary merger as a representative example for all considered configurations.

A major source of contamination of the emitted gravitational waves originates from spurious perturbations introduced by the initial data. As shown in detail in Ref.~\cite{Siemonsen:2023age}, even small amplitude long-lived oscillations induced in each star can lead to the emission of spurious radiation. The prompt response of the binaries, is particularly prone to these contaminations, since its amplitude scales linearly in the binary's mass-ratio (and $q=78$ in the most relativistic considered binary). The spurious gravitational wave emissions themselves scale as $\sim D^{-1}$ with the initial coordinate separation. Therefore, increasing the initial binary separation (in addition to the conformal rescaling as outlined above, and with details in Ref.~\cite{Siemonsen:2023age}) reduces spurious gravitational wave contaminations. This is the primary reason for considering such a large initial separation of $D/M_0=160$. Note, due to the scalar field's time dependence, comparing the gravitational radiation from the merger of identical binary boson stars with varying initial separation is non-trivial. Scalar interactions, active during the merger and ringdown, are sensitive to the phase-offset of the boson stars at the point of contact, which in general depends on the initial separation. We leave a study of these effects to future work, and hence, focus entirely on the separation $D/M_0=160$.

\begin{figure}
\includegraphics[width=0.48\textwidth]{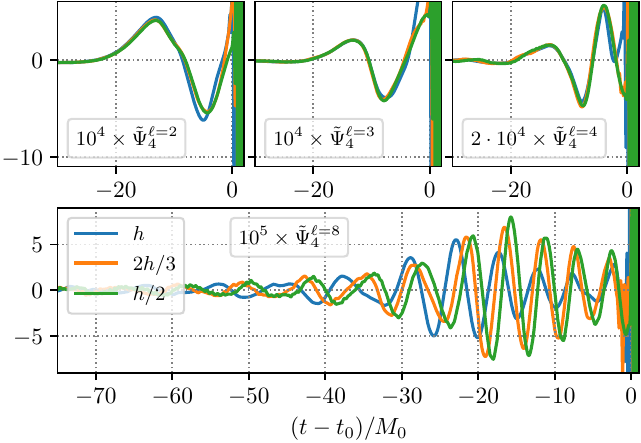}
\caption{The convergence behavior of the spin-weighted spherical harmonic components $\tilde{\Psi}_4^\ell=r_{\rm extr.}M_0\Psi_4^\ell$, extracted at a coordinate sphere of radius $r_{\rm extr.}/M_0=100$, with grid spacing $h$, corresponding to the prompt response of the $C=0.38$ binary boson star. All polar components are aligned, such that the first peaks of the $\ell=2$ modes align in time (as before, $t_0$ indicates the end time of the prompt response).}
\label{fig:promptconv}
\end{figure}

The gravitational waveforms are obtained by evaluating the Newman-Penrose scalar $\Psi_4$ on coordinate spheres with radii $r_{\rm extr.}/M_0\in \{ 50,76,100 \}$. We check explicitly that finite extraction radius effects are smaller than the uncertainty due to the discretization of the evolution grid. To quantify the latter, we compare the waveforms of the $C=0.38$ binary boson star merger at three different resolutions at $r_{\rm extr.}/M_0=100$. To that end, in \figurename{ \ref{fig:promptconv}} we show the convergence behavior of the prompt response of the remnant produced in this binary merger with increasing resolution. While the lowest resolution differs significantly, both the medium and high resolution waveforms align relatively well for all polar modes. In the case of $\ell=8$, there is a relative timeshift of the waveform of $\approx 0.6M_0$ between the medium and high resolutions. The uncertainties of waveforms shown in the main text are obtained from a time-averaged difference between the medium and high resolution waveforms shown in \figurename{ \ref{fig:promptconv}}.

\begin{figure}
\includegraphics[width=0.48\textwidth]{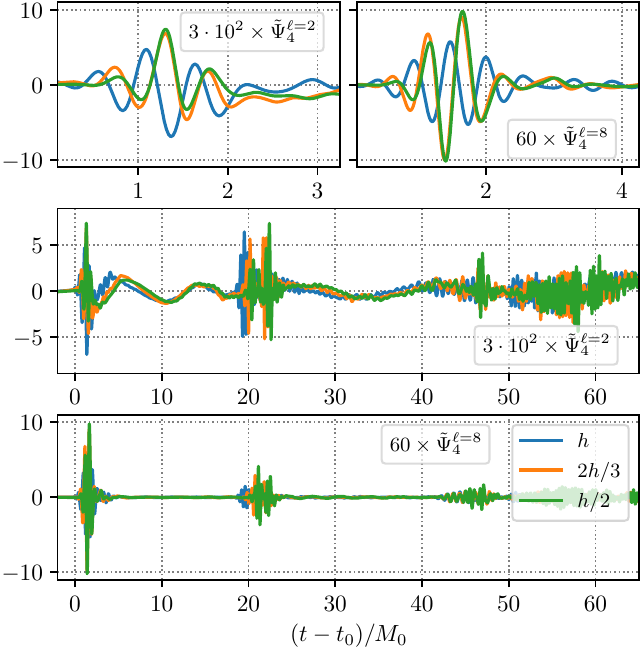}
\caption{As in \figurename{ \ref{fig:promptconv}}, however, now focusing on the convergence of the waveform emitted after the prompt ringdown of the $C=0.38$ remnant.}
\label{fig:burstconv}
\end{figure}

In \figurename{ \ref{fig:burstconv}}, we show the convergence behavior of the gravitational waveforms after the prompt response. Analogously to the discussion above, the lowest resolution waveform deviates significantly, while both the medium and high resolution gravitational waveforms align reasonably well. Among others, the burst's frequency, the burst period, burst amplitude, the amplitude of the long-lived gravitational wave component, as well as the frequencies of this long-lived component are reasonably converged in both the $\ell=2$ and $\ell=8$ polar modes. Naively, one may worry that the bursts in the waveform originate (at least partially) from the reflection of outgoing waves off AMR boundaries. However, we compared the burst waveforms at different extraction radii finding good agreement of $\tau_{\rm burst}$ (i.e., if bursts originated from such reflections, then $\tau_{\rm burst}$ would strongly depend on the extraction radius). The first burst's frequency $f_{\rm burst}$ is obtained from the centers of the peak of the Fourier transform of the $\ell=8$ waveform shown in the main text for $C=0.38$ (though the same result is obtained from a fit to $\ell=2$). The burst period $\tau_{\rm burst}$ is the time $t$ between the peak of each burst (if there are more than three bursts, we use the average of the times between bursts). The star's oscillation timescale $\tau_{\rm osc.}$ is obtained from fitting $\sin(at+b)+c$ to post-merger oscillations of $\max|\Phi|$ (as shown in \figurename{ \ref{fig:conv}}) for all four binaries considered in this work. Finally, the frequency $f_{\rm low}$ is obtained directly from Fourier transform of the type shown in the main text (for $C=0.28$ and $C=0.32$, the uncertainties of this method increase compared with the more compact binaries). Overall, we estimate the uncertainties of these timescales to be at the $\lesssim 10\%$ level. The uncertainties of the computed total radiated energy in gravitational waves we estimate to be $\approx 20\%$.

\subsection{Geodesic integrator}

To determine the light crossing times along various classes of null geodesics in the stationary boson star spacetimes, we solve the geodesic equations numerically. The latter equation is given by
\begin{align}
\frac{dx^\mu}{d\lambda}=u^\mu, & & \frac{du^\mu}{d\lambda}=-\Gamma^\mu_{\nu\gamma}u^\nu u^\gamma,
\label{eq:geodesic}
\end{align}
with Christoffel symbol $\Gamma^\mu_{\nu\gamma}$ associated with the boson star spacetime $g_{\mu\nu}$. Following from the stationarity and axisymmetry of $g_{\mu\nu}$, we have defined the conserved energy $E$ and angular momentum $L_z$ of the geodesics above. Together with the normalization condition $u_\mu u^\mu=0$, these are conserved along the geodesic. Initial conditions are provided by choosing an initial location $x^i(0)$ and spatial velocity $u^i(0)$. The temporal component $u^0(0)$ is then obtained from the normalization condition, which fixes the parameterization. The geodesic equations \eqref{eq:geodesic} are solved numerically using a fourth-order accurate Runge-Kutte method. We find it convenient to perform the integration on a Cartesian spatial grid, applying the Euclidean spherical to Cartesian coordinate transformations to the Lewis-Papapetrou coordinates of \eqref{eq:metricansatz}. To determine $\Gamma^\mu_{\nu\gamma}$, we use fifth-order accurate finite differences for the metric derivatives in the $r$ and $\theta$ directions, and employ linear order interpolations to obtain values of $f,l,g$ and $\Omega$ at an arbitrary location in $(r,\theta)\in(0,\infty)\times(0,\pi)$. During the integration of \eqref{eq:geodesic}, we monitor the constancy of $E$ and $L_z$ and assess the convergence based on the violations of the constraint $u^\mu u_\mu=0$ along the geodesic. The convergence of the latter is entirely limited by the resolution of the background spacetime (with grid spacing $\Delta r$ and $\Delta \theta$), as the typical stepsize of the Runge-Kutta integrator satisfies $\Delta\lambda\ll \Delta r, \Delta \theta$.

\end{document}